\documentstyle[aps,floats,graphicx]{revtex}

\begin{document}

\newcommand{\bec}{\begin{center}}
\newcommand{\ec}{\end{center}}
\newcommand{\be}{\begin{equation}}
\newcommand{\ee}{\end{equation}}
\newcommand{\beqn}{\begin{eqnarray}}
\newcommand{\eeqn}{\end{eqnarray}}
\newcommand{\bet}{\begin{table}}
\newcommand{\ent}{\end{table}}
\newcommand{\bib}{\bibitem}

\wideabs{

\title{
Interface-anisotropy induced asymmetry of intermixing in bilayers 
}

\author{P. S\"ule, M. Menyh\'ard, L. K\'otis, J. L\'ab\'ar, W. F. Egelhoff Jr.$^{\star}$} 
  \address{Research Institute for Technical Physics and Material Science,\\
Konkoly Thege u. 29-33, Budapest, Hungary,sule@mfa.kfki.hu,www.mfa.kfki.hu/$\sim$sule,\\
$^{\star}$ National Institute of Standards \& Technology, Gaithersburg, Maryland 20899\\
}
%\email{sule@mfa.kfki.hu}

\date{\today}

\maketitle

\begin{abstract}
The ion-sputtering induced intermixing is studied by molecular dynamics (MD) simulations
and by Auger electron spectroscopy depth profiling (AES-DP) analysis in Pt/Ti/Si substrate (Pt/Ti) and Ta/Ti/Pt/Si substrate (Ti/Pt) multilayers.
Experimental evidence is found for
the asymmetry of intermixing in Pt/Ti and in Ti/Pt. 
An unexpected enhancement of the injection of the heavy Pt atoms into the Ti substrate is observed
both by AES-DP and by MD simulations.
In Ti/Pt we get a much weaker interdiffusion than in Pt/Ti.
The asymmetry is
explained by the 
backscattering of hyperthermal particles at the mass-anisotropic interface and
which is reproduced by computer atomistic simulations.
The AES-DP measurements support our earlier predictions
(P. S\"ule, M. Menyh\'ard, Phys. Rev., {\bf B71}, 113413 (2005)) obtained for mass-anisotropic
bilayers.

%{\scriptsize {\em Keywords:} intermixing, interdiffusion, interface, film/substrate, Auger depth profiling, sputtering, computer simulations, thin film, multilayer, ion-solid interaction, molecular dynamics, Ti/Pt
%}
\end{abstract}
}

 There has been intense interest in the past few years in maintaining well-defined conditions
for the fabrication of ultra-thin film devices \cite{Schukin,Ramana,Egelhoff,Gomez05}.
  Controllable fabrication of interfaces is still a great challange for
nanoscience and nanotechnology \cite{Schukin,Egelhoff,Gomez05}.
It is necessary to understand the driving force of intermixing
in order to avoid or suppress interdiffusion during thin film growth.
Unfortunately our knowledge on the mechanism of interfacial mixing is
far from being complete \cite{Cai,Buchanan,Gomez}.

 Despite the apparent technological importance of thin film multilayers, the research of the atomic transport phenomena at 
interfaces, however, is less intense.
Auger electron spectroscopy (AES) depth profiling analysis has been used for the study of the sharpness and broadening
of interfaces during ion-sputtering \cite{Hofmann} in combination with transmission electron microscopy (TEM) \cite{Gnaser,Menyhard}.
 Computer simulations have revealed that
mass-anisotropy of the bilayer governs interdiffusion at the interface and
greatly influences surface morphology development during ion-sputtering
\cite{Sule_PRB05,Sule_SUCI}.

  In this Letter we present results which can be taken as the first direct experimental
evidence of mass effect on interfacial mixing.
We reproduce and explain the experimentally found asymmetry of intermixing by MD simulations.

%\section{The setup of the simulation}

 Classical constant volume molecular dynamics simulations were used to simulate the ion-solid interaction
using the PARCAS code \cite{Nordlund_ref}.
The computer animations can be seen in our web page \cite{web}.
Further details  are given in \cite{Nordlund_ref} and details specific to the current system in recent
communications \cite{Sule_PRB05,Sule_NIMB04}.
We irradiate the bilayers Pt/Ti and Ti/Pt (9 monolayers, ML film/substrate)
with 0.5 keV Ar$^+$ ions repeatedly with a time interval of 10-20 ps between each of
the ion-impacts at 300 K
which we find
sufficiently long time for the termination of interdiffusion, such
as sputtering induced intermixing (ion-beam mixing) \cite{Sule_NIMB04}.
 The initial velocity direction of the
impacting atom was $10$ degrees with respect to the surface of the crystal (grazing angle of incidence)
to avoid channeling directions and to simulate the conditions applied during ion-sputtering. 
We randomly varied the impact position and the azimuth angle $\phi$.
In order to approach the real sputtering limit a large number of ion irradiation are
employed using automatized simulations conducted subsequently together with analyzing
the history files (movie files) in each irradiation steps.
In this article we present results up to 100 ion irradiation which we find suitable for
comparing with low to medium fluence experiments. 100 ions are randomly distributed
over a $50 \times 50$ \hbox{\AA}$^2$ area which corresponds to $< 10^{15}$
ion/cm$^2$ ion fluence
and the removal of few MLs.
The size of the simulation cell is $110 \times 110 \times 90$ $\hbox{\AA}^3$ including
57000 atoms.
In order to reach the most efficient ion energy deposition at the interface,
we also initialize recoils placing the ion above the interface by $10$ \hbox{\AA} and
below the free surface in the 9 ML thick film. 
In this way  
we can concentrate directly on the intermixing phenomenon avoiding
many other processes occur at the surface (surface roughening, sputter erosion, ion-induced surface diffusion, cluster ejection, etc.) which weaken energy deposition at the interface.
Further simplification is that chanelling recoils are left to leave the cell
and in the next step these energetic and sputtered particles are deleted.

  The experimental setup is the following:
According to the XTEM results the thickness of the layers in samples are: Pt 160 \hbox{\AA}/Ti 130 \hbox{\AA}/Si substrate, and Ta 260 \hbox{\AA} (cap layer to prevent oxidation of Ti)/Ti 150 \hbox{\AA}/Pt 140 \hbox{\AA}/Si substrate. The interfaces are sharp and straight. 
For the sake fo simplicity we consider our multilayer samples as bilayers and we study the atomic transport processes at the Ti/Pt and Pt/Ti interfaces. 
We see no asymmetry in the interface widths of Ti/Pt and Pt/Ti bilayers (which includes roughness and mixing) using XTEM in Ta/Ti/Pt/Si (Ti/Pt) and in Pt/Ti/Si (Pt/Ti) in the as grown samples.
Both samples were AES depth profiled by applying sputtering conditions as following: ion energy $500-1500$ eV, projectile Ar$^+$, grazing angle of incidence $5^{\circ}-10^{\circ}$ degree (with respect to the surface), sample was rotated during ion bombardment.  
The atomic concentrations of Pt, Ti, Ta and Si were calculated by the relative sensitivity method taking the pure material's values from the spectra. The oxygen atomic concentration was calculated by normalizing the measured oxygen Auger peak-to-peak amplitude to TiO$_2$ \cite{Vergara}. The depth scale was determined by assuming that the sputtering yield ($Y_i$) in the mixed layer is the weighted sum of the elemental sputtering yields ($\sum_i X_i Y_i$). 
Two typical AES depth profiles recorded on samples of Pt/Ti and Ti/Pt at $500$ eV ion energy, are shown, respectively in Fig. 1. It is clear without any detailed evaluation that the Pt/Ti and Ti/Pt transitions are very different. While the Ti/Pt transition is a "normal" one, in case of Pt/Ti an unusual deep penetration of Pt to the Ti phase is observed. 
The interface widths between pure materials can be well characterized by the distance of points on the depth profile showing concentrations of $\sim 84$ \% and $\sim 16$ \% (step function true profile) if the transitions can be approximated by an error function. It can be supposed that the intrinsic surface roughness, $\sigma_{rough}$, is independent of ion bombardment induced mixing and thus the measured width, $\sigma_{meas}$, can be calculated as, $\sigma=\sqrt{\sigma_{meas}^2-\sigma_{rough}^2}$. Now $\sigma$ is the width as a result of the ion bombardment induced processes (mixing) and the finite inelastic mean free path of the signal electrons. The metal/substrate system interface was found to be sharp, thus correction is not necessary, while roughness of the other interfaces are about 15 \hbox{\AA}.  Though the Pt/Ti transition is far from being an error function type one, we still used the definition for comparative purpose.

In Fig. 1 we also recognize the presence of oxygen in the Ti layer in the Ti/Pt and metal/substrate interfaces. The later is the consequence of using Si substrate with native oxide. Since the bulk level of oxygen slightly correlated with the ion current intensity (the larger the ion current intensity the lower the oxygen AES signal) part of the oxygen is contamination occurring during the AES depth profiling process. The concentrations of the oxygen in the Ti/Pt interface neither correlate with the bulk level nor with the ion current intensity. Consequently we might conclude that the origin of the oxygen on the interface is due to the sample preparation process.
%------------------------------------------------------
\begin{figure}[hbtp]
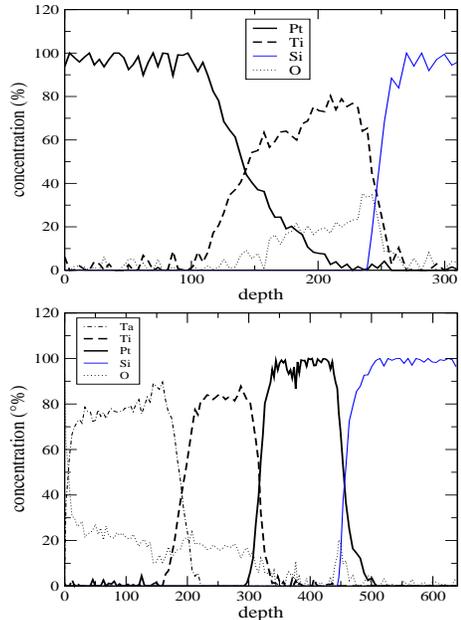

\begin{center}
\includegraphics*[height=4cm,width=6cm,angle=0.]{fig1a.eps}
\includegraphics*[height=4.2cm,width=6cm,angle=0.]{fig1b.eps}
\caption[]{
The concentration depth profile as a function of the removed layer thickness (\hbox{\AA}) obtained by AES depth profiling analysis
using ion-sputtering at 500 eV ion energy in Pt/Ti/Si substrate (upper Fig: 1a) and Ta/Ti/Pt/Si substrate (lower Fig: 1b).
}
\label{fig1}
\end{center}
\end{figure}
%------------------------------------------------------
Moreover the interface broadening does not correlate with the concentration of the oxygen in the Ti/Pt interface. 
Thus we conclude
that the atomic transport is not affected by the presence of the oxygen.

  The MD simulations of ion-sputtering and AES-DP reveal rather similar values for broadening
at the interface as well as reproduce the asymmetry of intermixing shown in Figs 1a and 1b.
In Fig 2 the evolution of the sum of the square atomic displacements of all intermixing atoms $\langle R^2 \rangle= \sum_i^N [{\bf r_i}(t)-{\bf r_i}(t=0)]^2$ (${\bf r_i}(t)$ is the position vector of atom 'i' at time $t$), can be followed as a function of the simulated ion fluence, and calculated as it has been given in ref. 
\cite{Sule_PRB05}. 
A result for the Cu/Co system is also given
to compare the magnitudes of intermixing in various bilayers and to
demonstrate the performance of the MD simulations.
The magnitude of ion-beam intermixing can also be characterized by the measurable quantity mixing efficiency $\xi$ \cite{Sule_PRB05,AverbackRubia} which is given as follows:
$\xi=\frac{\sigma^2}{\Phi F_D}$, where $\Phi$, and  $F_D$,
are the ion fluence (ion/$\hbox{\AA}^2$) and the deposited nuclear energy at the interface (eV/$\hbox{\AA}$), respectively.

  We get mixing efficiency value of $\xi \approx 10$ $\hbox{\AA}^5/eV$ which is
comparable with our experimentally measured value of $\sim 3-6$ $\hbox{\AA}^5/eV$ for Cu/Co.
The asymmetry of mixing can clearly be seen when $\langle R^2 \rangle (\Phi)$ and the depth profiles are compared in Ti/Pt and in Pt/Ti in Figs 1a,1b and 2. 
%------------------------------------------------------
\begin{figure}[hbtp]
\begin{center}
\includegraphics*[height=5.cm,width=6.5cm,angle=0.]{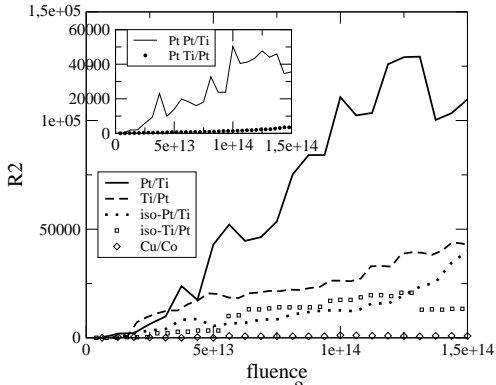}
\caption[]{
The simulated $\langle R^2 \rangle$ ($\hbox{\AA}^2$) in Pt/Ti, Ti/Pt and in Cu/Co as a function of the
ion-fluence (ions/cm$^2$) obtained during the ion-sputtering of these bilayers at 500 eV ion energy.
The dotted line (iso Pt/Ti) denote the results obtained for the artificial mass-isotropic
Pt/Ti bilayer. 
{\em Inset}: 
The simulated $\langle R^2 \rangle$ for Pt in Pt/Ti and in Ti/Pt
at 500 eV ion energy.
}
\label{fig2}
\end{center}
\end{figure}
%------------------------------------------------------
The computer animations of the simulations also reveal the stronger 
interdiffusion in Pt/Ti \cite{web}.
Moreover, in the inset in Fig 2 the higher mobility ($\propto$ $\langle R^2 \rangle$) of the energetic Pt atoms can be seen in
Pt/Ti.

  A relatively weak intermixing is found in Ti/Pt (the corrected ion-induced AES-DP interface width $\sigma \approx 13 \hbox{\AA}$ at 500 eV ion energy) and
a much stronger interdiffusion occurs in the Pt/Ti bilayer ($\sigma \approx 70 \hbox{\AA}$).
MD simulations provide $\sim 4$ ($\sim 10 \hbox{\AA})$ and $\sim 8$ ($\sim 20 \hbox{\AA})$ ML thick interface after 100 ion impacts, respectively.
As mentioned before, in our immiscible reference system, in Cu/Co we find also a weak mixing.

  Asymmetric AES depth profiles (when the broadening of A/B interface is different from that of B/A) have already been observed \cite{Hofmann,Barna}. The asymmetric behavior was explained by the large relative sputtering yield of the elements (preferential sputtering) causing different atomic mixing \cite{Hofmann} and/or enhanced bombardment induced interface roughening.
The asymmetry of intermixing is also known during thin film growth
\cite{Buchanan,MM}.
In the present experiment the relative sputtering yield of
$Y_{Pt}/Y_{Ti} \approx 0.7-0.9$ at 500-1500 eV
therefore the mechanism is different.
This kind of an ion-bombardment induced asymmetric mixing in metallic diffusion couples has not reported yet
at best of our knowledge.

  In recent papers we explained single-ion impact induced intermixing as an interdiffusion
process governed by the mass-anisotropy parameter (mass ratio) in these bilayers
\cite{Sule_PRB05,Sule_SUCI}.
Following this way of reasoning Ti/Pt (Pt/Ti) is a mass-anisotropic, while Cu/Co is
a mass-isotropic bilayer.
  It has already been shown in refs. \cite{Sule_PRB05,Sule_NIMB04} that the backscattering of the hyperthermal particles (BHP)
at the heavy interface leads to the increase in the energy density of the collisional cascade. 
We have found that the jumping rate of atoms through the interface is seriously affected by the
mass-anisotropy of the interface when energetic atoms (hyperthermal particles) are present and which leads to the
preferential intermixing of Pt to Ti \cite{Sule_PRB05,Sule_NIMB04}.
In this article we would like to explain the mixing asymmetry seen in the case of Pt/Ti and Ti/Pt using
the extension of this model.

The schematic view of the hyperthermic backscattering mechanism is shown in Fig 3.
  According to the BHP model, 
  in Pt/Ti, the energetic Ti light particles are backscattered downwards and confined below the heavy interface
in the bulk which results in increasing energy density below the interface.
%------------------------------------------------------
\begin{figure}[hbtp]
\begin{center}
\includegraphics*[height=3.5cm,width=6cm,angle=0.]{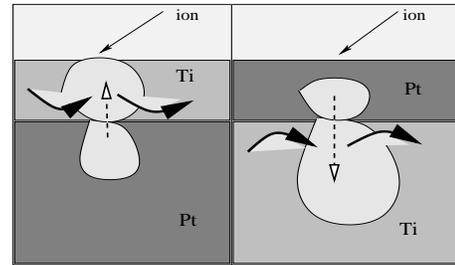}
\caption[]{
The schematic view of the hyperthermic atomic backscattering
mechanism in anisotropic bilayers.
The closed light regions represent the cascade region occurs at each
ion impact. Note that the lower lobe of the cascade is larger
in Pt/Ti than the upper lobe in Ti/Pt due to the downward scattering and the confinement of the light particles
indicated by the curved arrows.
The arrows with dashed lines show the atomic transport of Pt.
The path of the inpinging ions from the vacuum towards the free surface of the film/substrate system is also shown by arrows.
}
\label{fig3}
\end{center}
\end{figure}
%------------------------------------------------------
The increased lifetime and energy density of the collisional cascade or heat spike region below the interface
in Pt/Ti
leads to an enhanced injection of Pt atoms into the bulk and to a rather deep depth distribution of Pt shown in Fig 1a. We find a $18-22$ $\hbox{\AA}$ ($6-8$ ML) deep penetration of Pt into the Ti phase (and $36-44$ $\hbox{\AA}$ interface width) at $\Phi \approx 10^{15}$ ion/cm$^2$ fluence by MD simulations. Although the experimental $\sigma \approx 70 \hbox{\AA}$ value is still far from our simulated value, however, we expect that the penetration depth (intermixing length) of Pt could be increased by further ion-impacts.  
Nevertheless, MD simulations reproduce the mixing asymmetry and
we are able to explain the essential features of the phenomenon.
In Ti/Pt the Ti atoms are backscattered in the
overlayer film and ejected towards the surface. This leads to the reduction of the energy density
in the collisional cascade and hence to a weaker intermixing when
compared with the Pt/Ti system.
The BHP is the most effective when the projected range of the deposited energy is positioned
right at the interface.

 In order to clarify the details of intermixing,
  simulations have been run with mass ratio $\delta$ is artificially set to $\delta
 \approx 1$ (mass-isotropic), and we find that $\langle R^2 \rangle$ drops below the corresponding curve of Ti/Pt in Pt/Ti (see
Fig 2, dotted line, and animation \cite{web}).
We find it as a direct evidence of mass effect on sputtering induced intermixing
and on the observed asymmetry.
Hence, we expect that the energetic particle scattering mechanism is weakened
heavily
when
mass-anisotropy
vanishes and the magnitude of intermixing is also weakened approaching
%------------------------------------------------------
\begin{figure}[hbtp]
\begin{center}
\includegraphics*[height=4.5cm,width=6cm,angle=0.]{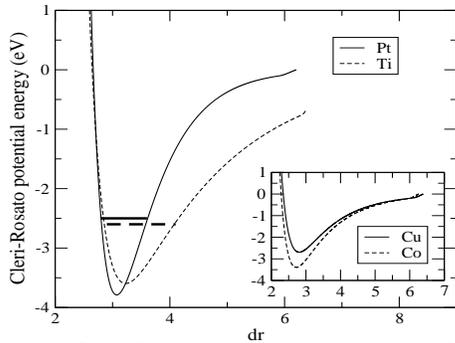}
\caption[]{
The Cleri-Rosato potential energy as a function of the
interatomic distance ($\hbox{\AA}$) in Ti and in Pt averaged for
an atom with its 12 first neighbors in the hcp and fcc lattices.
}
\label{fig4}
\end{center}
\end{figure}
%------------------------------------------------------
the $\langle R^2 \rangle$ values of Ti/Pt.
Therefore the observed mixing asymmetry can solely be explained by the
hyperthermic light particle backscattering mechanism at the mass-anisotropic interface.
The stronger simulated mixing in Ti/Pt than in Cu/Co
could mostly also be accounted for by mass effect.
The artificially set mass-isotropy in Ti/Pt results in the decrease of  
$\langle R^2 \rangle$ (filled diamonds in Fig. 2) which drops to the regime of Cu/Co. 
%In Ti/Pt, the mass effect explains
%partly the increase of intermixing rate, when compared with the low mixing rate
%of Cu/Co. 
In Ti/Pt, the heavy atoms at the interface block the transport path of the film atoms and reverse the flux of energetic Ti atoms.
In the artificially mass-isotropic Pt/Ti (iso-Pt/Ti), however, 
$\langle R^2 \rangle$ is similar to that of in iso-Ti/Pt and somewhat larger than in Cu/Co (see Fig 2)
hence mass effect does not solely explain intermixing in
these bilayers.

  To understand the physical origin of the higher intermixing in iso-Pt/Ti
and in iso-Ti/Pt than in Cu/Co,
we explain in general intermixing by a more complex anisotropy of the interface
then previously
expected. 
We can rule out the effect of thermochemistry (e.g. heat of mixing $\Delta H$) 
because
it has also been shown recently, that $\Delta H$ has no
apparent effect on intermixing during single-ion impact in this bilayer \cite{Sule_NIMB04}.
 Also, thermochemistry could not explain the mixing asymmetry observed in Ti/Pt and
in Pt/Ti since $\Delta H$ must be the same in magnitude in the two systems.
Moreover, simulated ion-sputtering with $\Delta H \gg 0$ (switching off the attractive term of the cross-potential
and using purely repulsive (Born-Mayer) Ti-Pt interaction potential for Pt/Ti)
leads to intermixing instead of the expected sharp interface (see also animations \cite{web}).
We reach the conclusion that there must be another anisotropy in the system.

   In order to find another anisotropy paramater,
  the employed interaction potential energy functions are plotted
in Fig ~(\ref{fig4}) as a function of the interatomic distance for Ti and for Pt. It can clearly be seen that
the thermal atomic size (the width of the potential valley at higher atomic kinetic energy) is different.
The atomic radius (the width of the vibration around the equilibrium lattice site) of the Ti atoms
is larger than that of in Pt. The atomic size anisotropy is even more robust at 1-2 eV energies above the energy minimum (hyperthermal atomic energies, see the horizontal lines in Fig 4). This explains the fact that the smaller and heavier Pt atoms are the first 
ballistic diffuser during ion-beam mixing \cite{Sule_NIMB04}. 
The atomic size difference (with smaller contribution) and the mass-anisotropy together could explain the stronger simulated (and measured) intermixing in Pt/Ti and in Ti/Pt than 
in Cu/Co. In the latter mass-isotropic bilayer no serious atomic size difference (anisotropy) is found (see the inset Fig ~(\ref{fig4})).
The mixing in the iso-Pt/Ti could also be understood by the size anisotropy:
the larger energetic Ti atoms are reversed at the interface and confined in the bulk.
The smaller energetic Pt atoms, however, are injected to the Ti bulk during
the cascade period.
Therefore the BHP mechanism is governed by a complex anisotropy of the
interface system: both by mass and size anisotropy.
In the iso-Ti/Pt the energy density of the cascade is weaker due to the
proximity of the free surface hence the influence of the size difference
is less efficient on intermixing.

 We could explain the observed mixing asymmetry between Ti/Pt and Pt/Ti in terms
of the backscattering of energetic particles at the mass-anisotropic interface.
The mass effect might not solely explain intermixing in Pt/Ti:
the atomic size and mass-anisotropy might together be responsible for
the observed strong interfacial mixing and
the weaker mixing in Cu/Co is explained by atomic size and mass-isotropy.

%\section{acknowledgement}
{\scriptsize
This work is supported by the OTKA grants F037710
and T043704 
from the Hungarian Academy of Sciences.
We wish to thank to K. Nordlund 
for helpful discussions and constant help.
The work has been performed partly under the project
HPC-EUROPA (RII3-CT-2003-506079) with the support of
the European Community using the supercomputing 
facility at CINECA in Bologna.
The help of the NKFP project of 
3A/071/2004 is also acknowledged.
}

\end{document}